\documentclass[12pt]{article}

\usepackage{scicite}

\usepackage{times}

\newcounter{lastnote}

\title{Unveiling the Formation of Massive Galaxies}

\author
{Christopher J. Conselice$^{1\ast}$\\
\\
\normalsize{$^{1}$California Institute of Technology}\\
\\
\normalsize{$^\ast$To whom correspondence should be addressed; E-mail:  cc@astro.caltech.edu.}
}

\date{}

\begin{document}


\baselineskip24pt


\maketitle



One of the primary goals of modern astronomy is understanding the
formation history of galaxies.  This problem
can be subdivided into separate sub-questions, perhaps the most 
straightforward 
is understanding the formation of the most massive galaxies.  There
are well defined predictions for how massive galaxies should form [1,2]
and massive galaxies are the easiest to study since they
are usually the brightest at any given epoch.  Massive
galaxies in the nearby universe contain most of the stars [3] and are 
generally composed of old
stellar populations [4].  However, it is difficult if not 
impossible to determine, within a few billion years, the ages of
these galaxies using modern methods for dating stellar
populations [4].  The solution to understanding the formation of
massive galaxies is to study them in the early universe when they are
forming at high redshift.

There are two questions concerning the formation of massive galaxies
that astronomers are beginning to answer.  These are understanding when and 
how these systems formed.  Historically, most attention has been placed on 
determining
when massive galaxies formed by identifying and measuring properties 
of galaxies at high redshift.   The first ultraviolet 
selected samples of high redshift (at $z \sim 3$; about 11 billion years ago) 
galaxies demonstrated that these
systems have number densities and clustering properties similar
to nearby massive galaxies [5,6].  However, follow up
determinations of the stellar masses of these galaxies, or the amount
of mass in a stellar form, showed that these galaxies have lower stellar 
masses, similar
to the mass of the bulge of our own Milky Way galaxy [7].
Very few to none of these massive systems have stellar masses within a factor
of ten of the most massive galaxies in the modern universe [7,8]. 

The total integrated stellar mass density at these redshifts is  also
roughly a factor of ten lower than the stellar mass density today  
[9].  
It appears from these observations, and the fact that the star formation
rate within galaxies is high up until redshifts $z \sim 1$ [10], that
some massive galaxies did not form all their mass early. This is consistent 
with the Cold Dark Matter model for
structure formation that predicts the most massive objects
form gradually through accretion and merging [1].  Another possibility
is that there are galaxies at
redshifts $z \sim 3$ that are not identified in ultraviolet selected
redshift surveys because they are made up of old stars or contain large
amounts of dust.  Both situations create galaxies with red spectral
energy distributions which would be missed in traditional
ultraviolet selected surveys [11].

It has been argued that populations of these red, possibly old and massive,
galaxies have been found at $z \sim 1.5 - 3$ [11,12].  These
systems are characterized by rest-frame optical colors similar to colors of
nearby normal galaxies, and their clustering and stellar mass properties 
suggest that they are massive galaxies [13].  The integrated 
stellar mass density of these
galaxies is roughly similar to the stellar mass density of the
 star bursting population
at similar redshifts [11].  Determining how common these massive, possibly
evolved, 
galaxies are will require deeper and wider near infrared imaging and 
spectroscopic surveys that are just now feasible.

Massive galaxies may also not be easily visible or identifiable in optical or
near infrared surveys, because of high amounts of light extinction by dust.
In the last decade, a significant population of bright sub-mm galaxies were
found at redshifts $z > 2$ that are potentially precursors of contemporary 
massive galaxies [14].
These galaxies were discovered in deep sub-millimeter surveys that sample
rest-frame far infrared radiation which originates from dust grains heated by
photons from massive young stars. 
The dust in these galaxies absorbs energetic photons, and it is not
clear how much light from stars in these galaxies should be seen. However, 
the internal kinematics of these systems, based on
the velocity width of the CO emission line, suggests that they
are massive galaxies [15]. 
It is not yet known if these
systems represent a phase of evolution that relates to galaxies chosen in 
ultraviolet 
and near infrared selected samples. 

In addition to understanding when massive galaxies formed, astronomers
are also investigating how this formation occurred. Assuming that we are not
missing a large population of massive galaxies  at high redshift,  the
higher number density of these systems at lower redshifts 
suggests that massive galaxies must have formed gradually 
through time.   How does this occur?  There are several possibilities, 
including:
major mergers between galaxies of similar mass to build larger galaxies,
minor mergers of smaller satellites, and the accretion of intergalactic
gas which is converted to stars.  Understanding which of these modes is
responsible for forming massive galaxies is a fundamental problem that
is just now being addressed.

Perhaps the most popular explanation is that the most massive galaxies formed
through multiple major merger events.   Major galaxy mergers are in fact a 
prediction of the Cold Dark 
Matter cosmology, and are found to occur in simulations of
galaxy formation [1].  Understanding and tracing the extent of
major mergers in the early universe is however difficult.  
Recently it has been shown that by using high resolution Hubble Space 
Telescope imaging, it
is possible to determine the formation modes of galaxies.   Specifically, 
we can identify systems undergoing major mergers by their
peculiar and distorted structures.  Within the Hubble
Deep Field North the merger rate and history have been
traced in detail as a function of galaxy luminosity and stellar
mass [16].   Galaxies undergoing the most merging at
high redshift, $z > 2$,  are the most luminous and massive galaxies.  
By tracing the merger history for the most massive galaxies it appears that 
very few mergers occur in massive galaxies at lower redshifts 
[16].  This is consistent with finding massive
evolved galaxies at modest redshifts [12] and is in direct conflict
with the predictions of Cold Dark Matter models.  Based on these observations, 
it appears that massive galaxies do not form rapidly
early in the universe, as in the traditional early monolithic collapse
picture, but nor are they forming gradually throughout time as in Cold
Dark Matter simulations. 

It is however still not clear how the merging ultraviolet bright systems 
at $z \sim 2.5$ relate to the sub-millimeter and
near infrared selected galaxies found at similar redshifts.  It is likely that
these represent various phases of galaxy evolution whose time-scales
are still unknown.  It is also likely that the environment of galaxies
is a significant factor in their evolution [13] such that those in
denser areas are forming earlier than galaxies in lower density environments.
Little is understood of this effect at high redshift, but future 
deep infrared surveys should address this problem in the coming years.

\begin{quote}
{\bf References and Notes}

\begin{enumerate}
\item S. Cole., et al., {\it Monthly Notices of the Royal Astronomical Society\/}, 319, 168 (2000)
\item C. Cesare., G. Carraro, {\it Monthly Notices of the Royal Astronomical Society\/}, 335, 335 (2002)
\item M. Fukugita, C.J. Hogan, P.J.E. Peebles, {\it Astrophys. J.\/}, 503, 518 (1998)
\item G. Worthey, {\it Astrophys. J.\/}, 95, 107 (1994)
\item C. Steidel et al. {\it Astrophys. J.\/}, 462, L17 (1996)
\item M. Giavalisco, et al. {\it Astrophys. J.\/}, 503, 543 (1998) 
\item C. Papovich, M. Dickinson, H. Ferguson, {\it Astrophys. J.\/}, 559, 620 (2001)
\item A. Shapley, et al. {\it Astrophys. J.\/}, 562, 95 (2001)
\item M. Dickinson, et al. {\it Astrophys. J.\/}, 587, 25 (2003)
\item P. Madau, L. Pozzetti, M. Dickinson, {\it Astrophys. J.\/}, 498, 106 (1998)
\item M. Franx, et al. {\it Astrophys. J.\/}, 587, 79L (2003)
\item K. Glazebrook, et al. pre-print, astro-ph/0401037 (2004)
\item E. Daddi, et al. {\it Astrophys. J.\/}, 588, 40 (2003)
\item S. Chapman, et al. Nature, 422, 695 (2003)
\item R. Genzel, et al. {\it Astrophys. J.\/} 584, 633 (2003)
\item C. Conselice et al. {\it Astron. J.\/} 126, 1183 (2003)
\end{enumerate}
\end{quote}

\end{document}